\documentclass[a4paper]{aipproc}

\layoutstyle{6x9}

\usepackage{graphics}
\usepackage{psfrag}
\usepackage{epsfig}
\usepackage{psfig}
\usepackage{rotating}
\usepackage{epsf}

\newcommand{\jour}[4]{{#1}\textbf{{#2}} (#3) #4.}

\newcommand{\NPA}{Nucl.\ Phys.\ \textbf{A}}
\newcommand{\NPB}{Nucl.\ Phys.\ \textbf{B}}
\newcommand{\PLB}{Phys.\ Lett.\ \textbf{B}}

\newcommand{\beq}{\begin{equation}}
\newcommand{\eeq}{\end{equation}}
\newcommand{\beqa}{\begin{eqnarray}}
\newcommand{\eeqa}{\end{eqnarray}}

\newcommand{\bma}{\begin{array}{cc}}
\newcommand{\ema}{\end{array}}

\def\3{{\ss}}

\def\vek #1 {\overrightarrow {#1}}

\begin{document}

\title 
      [Chiral dynamics in few--nucleon systems]
      {Chiral dynamics in few--nucleon systems}

\classification{}
\keywords{Effective field theory; chiral perturbation theory; isospin violation}

\author{Evgeny Epelbaum$^\star$, Ulf-G. Mei{\ss}ner, Walter Gl\"ockle,\\
C. Elster, H. Kamada, A. Nogga, H. Witala}{
  address={$^\star$Forschungszentrum J\^ulich, Institut f\"ur Kernphysik (Th), D-52425 J\"ulich,
    Germany\\ Email: Evgeni.Epelbaum@tp2.ruhr-uni-bochum.de
}
}

\copyrightyear  {2001}

\begin{abstract}
We report on recent progress achieved in calculating various few--nucleon 
low--energy observables from effective field theory. Our discussion includes
scattering and bound states in the 2N, 3N and 4N systems and isospin 
violating effects in the 2N system. We also establish a link between 
the nucleon--nucleon potential derived in chiral effective field theory and various 
modern high--precision potentials.  
\end{abstract}

\date{\today}

\maketitle

\section{Introduction}

Chiral effective field theory (EFT) offers a systematic and controlled method
to study the dynamics of few--nucleon systems.  In the approach proposed
by Weinberg \cite{Weinb1,Weinb2}, one starts from an effective Lagrangian for nucleon and pion
fields as well as external sources, in harmony with chiral and gauge invariance.
The effective nucleon--nucleon (NN) 
potential is constructed using  a unitary transformation to avoid any energy-dependence
and applying systematic power counting, as described in ref. \cite{EGM1}.
To leading order, one has the one--pion exchange (OPE) together with two contact
terms accompanied by low--energy constants (LECs). At 
next--to--leading order (NLO), renormalizations of the OPE, the leading two--pion
exchange (TPE) diagrams and seven more contact operators appear. At NNLO, one has to include
the subleading TPE with one insertion of 
dimension two pion--nucleon vertices (the corresponding coupling constants are denoted by $c_{1,3,4}$).
The potential is then used in a properly regularized 
Lippmann--Schwinger equation (e.g. by a sharp or exponential
momentum cut-off) to generate the bound and scattering states,
as detailed in ref.\cite{EGM2}. The iteration of the potential leads to a non--perturbative
treatment of the pion exchange which is of major importance to properly describe
the NN tensor force. The nine LECs are to be determined from a fit to the
low $np$ partial waves. Here, we discuss implications of the uncertainties in the values of the $c_i$'s 
to various properties of chiral forces. Apart from the already published NNLO potential
with numerically large values of the $c_i$'s, taken from ref.\cite{Paul}, we construct 
the NNLO* version with (numerically) reduced and fine tuned values of $c_3$, $c_4$. We also discuss 
physical mechanisms, which can explain such smaller values of these low--energy constants. 
In contrast to the 
NNLO version, the NNLO* chiral potential does not lead to deeply lying bound states. This makes 
it more suitable for many--body applications. Both versions reproduce $np$ phase shifts in  most 
partial waves fairly well up to $E_{\rm lab} \sim 200$ MeV.
In the same framework, one can also include
charge symmetry breaking and charge dependence of the nuclear force 
by including the light quark
mass difference and elecromagnetic corrections \cite{WME}. This allows to predict differences between 
some phase shifts in $pp$ and $nn$ systems. 
The extension to three-- and
four--nucleon systems has also been started using the NLO potential \cite{ourPRL}.
At this order, no 3N forces (3NF) appear and one can make parameter--free predictions.
We also present some first NNLO* results including only  two--nucleon
forces.  We further show that the 
numerical values of the LECs can be understood on the basis of phenomenological
one--boson--exchange (OBE) models \cite{EGME}. We also extract these values from various modern
high accuracy NN potentials and demonstrate their consistency and
remarkable agreement with the values in the chiral effective field theory approach.
This paves a way for estimating the  low--energy constants of operators with 
more nucleon fields and/or external probes.

\section{Few--nucleon forces in chiral EFT}

Chiral EFT is a powerful tool, which allows to calculate low--energy observables performing
an expansion in powers of the low--energy scale $Q$ associated with momenta of external
particles.\footnote{To be more precise, $Q$ is associated with the 4--momenta of external pions
and 3--momenta of external nucleons.} The pion mass $M_\pi$ is treated on the same footing as $Q$.
To select the relevant diagrams contributing to 
the S--matrix at a certain order, one makes use of power counting. This consists of 
a set of rules to calculate
the power $\nu$ of the low--energy scale $Q$ for any given diagram. The precise meaning of
the power counting scheme depends on the system one is investigating. 
In what follows, we will concentrate on various low--energy processes between few (2, 3 and 4) nucleons.
The essential complication in that case compared to pion--pion or pion--nucleon scattering is given
by the fact, that the nucleon--nucleon interaction is nonperturbative even at very low energies.
To deal with this problem Weinberg proposed to use time--dependent (``old--fashioned'') perturbation theory 
instead of the covariant one. The expansion of the S--matrix, 
obtained using this formalism, has the form of a Lippmann--Schwinger
equation with an effective potential, defined as the sum of all
diagrams  without pure nucleonic intermediate states. Such states
would lead in "old--fashioned" perturbation 
theory  to energy denominators, which are by a factor of $Q / m$ 
smaller than those from the states with pions and which destroy 
the power counting. Here and below, $m$ denotes the nucleon mass. 
The effective potential is free from such small 
energy denominators and can, in principle, be calculated perturbatively to any given precision.
This strategy has been followed in the pioneering work by Ord\'o\~nez et al.~\cite{ord}.
The effective NN Hamiltonian derived in this formalism possesses some unpleasant properties: it depends, in general,
explicitly on the energy of incoming nucleons and  is not Hermitean. This complicates its application
to few--nucleon systems.
In order to avoid these problems we construct an effective NN Hamiltonian using the method of unitary
transformation (projection formalism), as described in ref.~\cite{EGM1}. For that we modify 
Weinberg's power counting in an appropriate way. All details are given in \cite{EGM1}. In what follows, 
we will show the qualitative and the quantitative
 results for the effective potential in the first few orders of the chiral expansion. 

\begin{figure}[htb]
\centerline{
\psfig{file=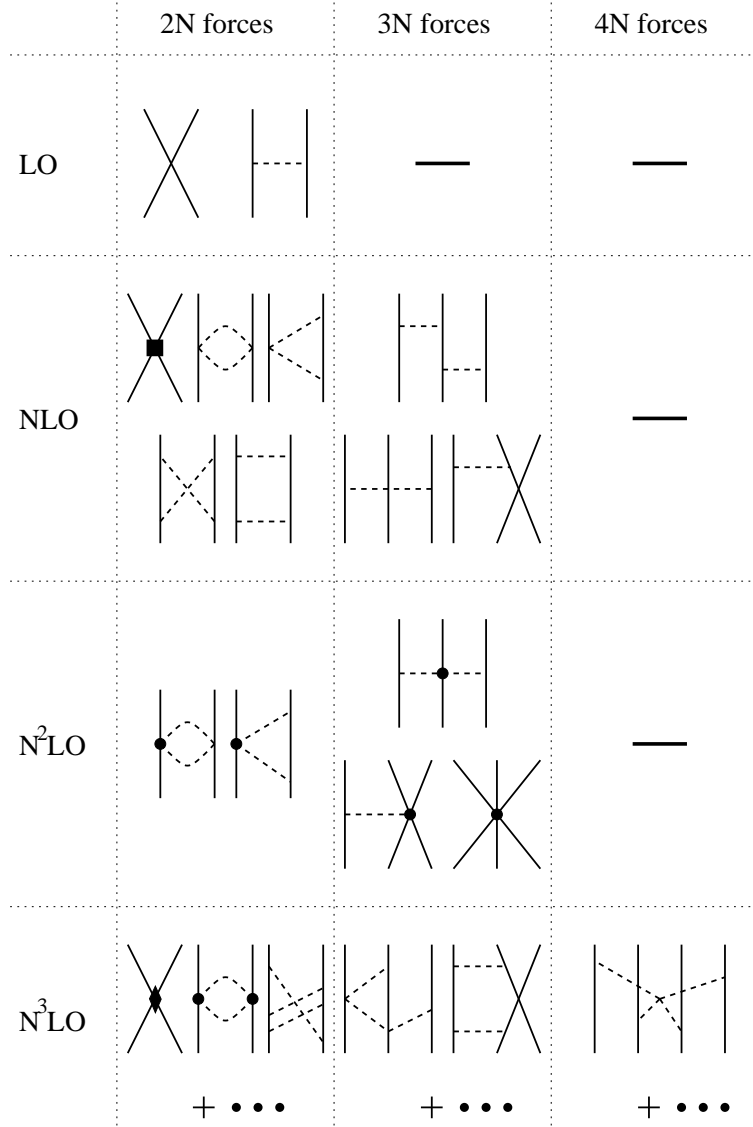,width=4in}}
\caption{\label{fig1}First orders in the chiral expansion for few--nucleon forces, as explained in the text. 
Solid (dashed) lines denote nucleons (pions).
The heavy dots, filled square and 
filled diamond denote the $\Delta_i =1$, $\Delta_i=2$ and $\Delta_i=4$ vertices, respectively.}
\end{figure}

Let us begin with the power counting. As pointed out above, any (time--ordered) diagram 
${\cal T}$ contributing to the 
few--nucleon scattering process scales as:
\begin{equation}
\label{nu}
{\cal T} \sim \left( Q /\Lambda_\chi \right)^\nu~,
\end{equation} 
where $\Lambda_\chi \sim 1$ GeV is the typical scale of chiral symmetry breaking.
For any diagram with $E_n$ nucleons, $L$ loops  and 
$V_i$ vertices of type $i$ one has (we only consider connected diagrams):
\begin{equation}
\label{powc}
\nu = -4 + 2 E_n + 2 L + \sum_i V_i \Delta_i\;,
\end{equation}
where each vertex carries the index $\Delta_i$ (also called chiral dimension) given by
\begin{equation}
\label{chirdim}
\Delta_i = d_i + \frac{1}{2} n_i - 2\;.
\end{equation}
Here, $n_i$ is the number of nucleon field operators and $d_i$ is the number of derivatives
(or pion mass insertions). Due to spontaneously broken chiral symmetry, the index $\Delta_i$ cannot
be negative. As a consequence, the power $\nu$ of the low--energy scale $Q$ is bounded from below 
and a systematic and perturbative expansion for an effective Hamiltonian becomes possible. 
Let us now concentrate on the first few orders in the chiral expansion.
\begin{itemize}
\item {\bf LO ($\nu=0$)}
The power $\nu$ of the low--energy scale $Q$ takes its minimal value $\nu =0$ for 
tree diagrams with two nucleons ($E_n=2$) and with all interactions of dimension
$\Delta_i=0$. Thus, the potential at LO is given by OPE and contact interactions without derivatives, see
fig.~\ref{fig1}.
There are no 3N and 4N forces at LO.
\item  {\bf NLO ($\nu=2$)}
The first corrections to the 2N potential are given by tree diagrams with one insertion of $\Delta_i = 2$
interaction (seven independent contact interactions with two derivatives) as well as by one--loop graphs
with all leading vertices\footnote{Only irreducible topologies have to be taken into account in 
``old--fashioned'' perturbation theory. On the contrary, in the projection formalism one has also to
include reducible topologies, as explained below.} 
(leading chiral TPE). Nominally, one has also three--nucleon forces given by tree
diagrams with $\Delta_i=0$ vertices. In the projection formalism it turnes out, that 
the total contribution from those diagrams vanishes. If ``old--fashioned'' perturbation theory is used to derive 
the effective potential, the contribution of the corresponding 3N 
force cancels against energy dependent part of once iterated 
LO potential \cite{Weinb1,kolck2}.
Thus, there are no 3N and 4N forces at this order.
\item {\bf N$^2$LO ($\nu=3$)}
The N$^2$LO 2N potential is given by the subleading chiral TPE with one $\pi \pi NN$ vertex of dimension
$\Delta_i=1$. There are three independent structures in the Lagrangian, which contribute to NN scattering.
The corresponding LECs are denoted by $c_{1,3,4}$ \cite{BKMrev}. 
Note that no additional contact interactions appear at
that order. There are first nonvanishing three--nucleon forces. 
\item {\bf N$^3$LO ($\nu=4$)}
The chiral potential at this order has not yet been worked out completly. Some calculations were performed by 
Kaiser \cite{Kaiser}. At this order one has first 4N forces as well as a large number of various 3N and 2N 
interactions (including two-- and three--pion exchange graphs).
\end{itemize}
Note that one can easily write down the contributions to the effective potential
within ``old--fashioned'' perturbation theory, which correspond to the diagrams shown in fig.~\ref{fig1}.
The contributions in the projection formalism are, however, not easily recoverable without knowledge 
of the explicit operator expressions. 
The diagrams of fig.~\ref{fig1} should therefore serve only as a guidance. 
For the precise numerical prefactors as well as the energy denominators
corresponding to the graphs shown the reader is
refered to ref.\cite{EGM1}. 
We would like to point out one of the most important qualitative findings of chiral EFT applied to 
few--nucleon systems \cite{Weinb2}, \cite{kolck2}. As shown in fig.~\ref{fig1}, 
the chiral power counting eq.~(\ref{powc}) suggests, that
3N forces are weaker than 2N ones, 4N forces are weaker than 3N ones etc..  

\section{Applications}
In what follows, we will discuss application of the described formalism to 2N, 3N and 4N systems,
consider some isospin violating effects and establish connection between chiral NN forces and OBE models.

\subsection{Two nucleons}

\begin{figure}[htb]
\centerline{
\psfig{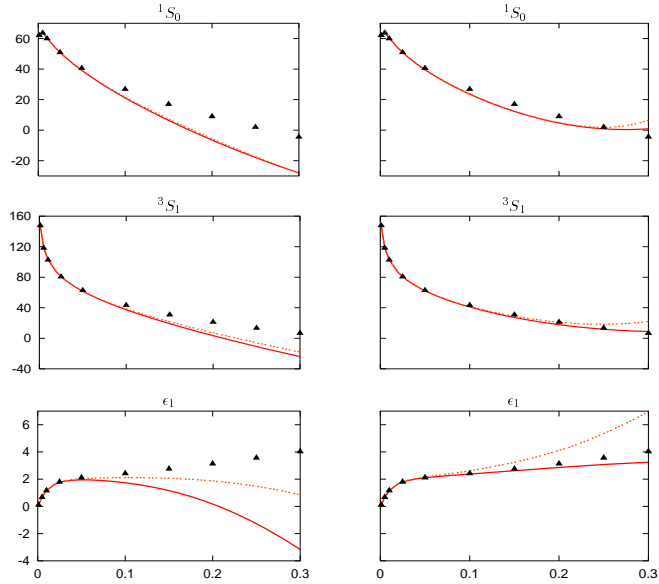}}
\vspace{-3cm}
\caption{\label{fig2}Phase shifts at NLO (left panel) and NNLO* (right
  panel) versus the lab energy (in GeV) in comparison to the Nijmegen PSA. 
  The solid (dashed) line corresponds to $\Lambda = 500$ ($\Lambda=600$) MeV.}
\end{figure}

Let us now concentrate on the parameters entering the NN potential. The two unknown LECs at LO associated
with the contact interactions without derivatives have to be fixed by a fit to S--wave phase shifts at low
energies. The leading chiral TPE at NLO is parameter--free. At this order one has in addition 
seven unknown LECs, which 
corresponds to contact terms with two derivatives. Those constants as well as the two LO LECs are fixed
from fit to phase shifts in S-- and P--waves and to the mixing angle $\epsilon_1$. Thus, one has nine unknown 
constants at this order, see ref.~\cite{EGM2} for more details.
As already stressed before, the subleading TPE at NNLO depends on the LECs $c_{1,3,4}$, which corresponds to 
$\pi \pi NN$ vertices of dimension $\Delta_i=1$. Precise numerical values of these constants are 
crucial for some properties of the effective potential. Clearly, the subleading $\pi \pi NN$ verices 
represent an important link between NN scattering and other processes, such as $\pi N$ scattering. Ideally, 
one would like, therefore, to take their values from the analysis of the $\pi N$ system. Various calculations
for $\pi N$ scattering have been performed and are published. From the $Q^2$ analyses \cite{BKM95} one gets:
$c_1 = -0.64\,,  c_3 = -3.90\,, c_4=2.25\;$.
Here  all values are given in GeV$^{-1}$.
{}From various $Q^3$ calculations \cite{BKM95}, \cite{BKM97}, \cite{Moi98}, \cite{FMS99}, \cite{Paul} one gets
the following bands for the $c_i$'s:
\begin{equation} 
\label{q3}
c_1 = -0.81 \ldots -1.53 \,, \quad \quad c_3 = -4.70 \ldots -6.19\,, \quad \quad c_4 = 3.25 \ldots 4.12\,.
\end{equation}
Recently, the results from the $Q^4$ analysis have become available \cite{FM00}.
At this order the S--matrix is sensitive to 14 LECs (including $c_{1,3,4}$), which 
have been fixed from a fit to $\pi N$ phase shifts. It turnes out that different
available phase shift analyses (PSA) from refs.~\cite{Koch86}, \cite{Mat97} and \cite{SAID} 
lead to sizable variation in the actual values of the LECs. In
particular, the dimension two LECs acquire a quark mass
renormalization. The corresponding shifts are proportional to
$M_\pi^2$. The renormalized $c_i$'s are denoted by  $\tilde{c}_i$. 
A typical fit  
based on the  phases of ref.~ \cite{Mat97} leads to:
\begin{equation}
\label{mats}
\tilde{c}_1 = -0.27 \pm 0.01\,, \quad \quad \tilde{c}_3 = -1.44 \pm 0.03\,, \quad \quad \tilde{c}_4=3.53
\pm 0.08 \;.
\end{equation} 
However, using the older Karlsruhe or the VPI phases as input, one
finds sizeable variation in the $\tilde{c}_i$.  Alternatively, one can
also keep the $c_i$ at their third order values and fit the fourth
order corrections separately, see \cite{FM00}. Due to the fitting
range chosen and uncertainties in the isoscalar amplitudes, these pieces are not very
well determined. In principle, these fits could be improved by
including the scattering lengths determined from pionic hydrogen/deuterium.
It should also be stressed that numbers consistent with the band given
in eq.(\ref{q3}) have been obtained in \cite{BL} using IR regularized
baryon chiral perturbation theory and dispersion relations. Comparable values 
of $c_{1,3,4}$ have also been obtained by Rentmeester et al.~from analysis of the $pp$ data 
\cite{Rent}.\footnote{Note, however, that it is not possible to fix all three constants in 
this prosess. For that reason the constant $c_1$ has been fixed at the value $c_1 =-0.77\,$GeV$^{-1}$.} 
A few comments are in order. First of all, the  numerical values of
some of the $c_i$'s appear 
to be quite large. Indeed, from the naive dimensional analysis one would expect, for example,  $c_3$ 
to scale like:
\begin{equation}
\label{scaling}
c_3 \sim {\ell} /2m \sim {\ell} / 2\Lambda_\chi \,,
\end{equation}
where ${\ell}$ is some number of order one. Taking the value $c_3 = - 4.70$ from ref.~\cite{Paul}
and $\Lambda = M_\rho = 770$ MeV we end up with ${\ell} \sim - 7.5$ . Such a large value 
can be partially explained by the fact that the $c_{3,4}$ 
are to a large extent saturated by the $\Delta$--excitation. This implies that the different and smaller 
scale, namely $m_\Delta - m \sim 293$ MeV, enters the values of these constants, see \cite{BKM97}.
More work on pion-nucleon scattering (dispersive versus chiral
representation), new dispersive analyses and more precise low-energy data
are needed to pin down these LECs to the precision required here.
Applied to  NN scattering, the large numerical values of the $c_i$'s might lead to a slow 
convergence of the low--momentum expansion. 
Another consequence of the large $c_i$'s for the NN system is the appearance of spurious deeply 
bound states in low partial waves, which can be traced back to a very strong attractive central 
potential related to the subleading TPE. The spurious states do not influence low--energy observables, as explained
in \cite{EGM2}. The important consequence is, however, that strong 3NF are needed\footnote{Calculated with 
only 2N forces, the triton is underbound by about 4 MeV \cite{EGMtoappear}.}. 
Note that despite this very different
scenario from what is expected in conventional nuclear physics (small 3N forces), the individual 
contributions of the 2N and 3N forces to observables can, in principle, not be observed experimentally.
Since the actual values of the $c_i$'s may possibly change in future analyses of the $\pi N$ system
(higher orders, more precise PSA, etc.), we constructed the NNLO* version of the NN potential, in which
we basically subtracted the $\Delta$--contributions to these LECs and allowed for some fine tuning. This 
results in numerically reduced values of the $c_{3,4}$: $c_3 = -1.15\,$GeV$^{-1}$, $c_4=1.20\,$GeV$^{-1}$. 
As a consequence, no unphysical deeply bound states appear 
This is partially motivated by the fact that the $\Delta$ is not included as an explicit degree of freedom 
in existing OBE models leading to very good quantitative description of observables. Some steps along a deeper
understanding of the surprisingly modest role of the $\Delta$ in the NN system within boson exchange
 models have been undertaken in the framework of the Bonn potential \cite{HME,Juelpirho}. It was pointed out 
that there are strong cancellations between the TPE, whose dominant part is given
by diagrams with intermediate $\Delta$--excitations, with the $\pi
\rho$--exchange. Such a cancellation should ultimately also be
observed in EFT studies, but that would require a consistent power
counting scheme including vector mesons. Such a scheme has not yet
been constructed. Let us finally point out that we do not know at the moment, 
whether the NNLO or NNLO* versions are closer to reality.
Further studies of different processes as well as going to higher orders in the NN system may shed some light 
on to this question. In what follows, we will use the NNLO* potential in our 3N and 4N calculations as well as 
for comparison with conventional models.\footnote{The application of the NNLO potential to few--nucleon systems
is technically more complicated.}
We are now in the position to give quantitative results.
In fig.\ref{fig2} we show the two $np$ S-wave  phase shift $^1S_0$ and
$^3S_1$ and the $^3S_1 - ^3D_1$ mixing parameter at NLO (left panel)  
and NNLO* (right panel) in comparison to
the Nijmegen PSA. To regularize the LS equation, we have used an
exponential regulator $f_R (\vec{p}) = \exp(-p^4/\Lambda^4$) (for details,
see \cite{EGM2}). The two lines correspond to cut-offs $\Lambda = 500$
and $600$~MeV.  We note that the description of the phases improves 
when going from NLO to NNLO* and that also the cut-off dependence gets
weaker (especially at low energies). This is to be expected from a converging EFT and we emphasize 
again that this is not the result of an increasing number of 
free parameters.  For the other
phase shifts and a more detailed discussion, see \cite{EGMtoappear}.

\subsection{Isospin violation}

It is well known that charge symmetry (CS) and charge independence (CI) of the
nuclear force are violated. In the Standard Model, these are
manifestations of isospin violation (IV). There are two distinct
sources of IV. In pure QCD, the light quark mass difference $m_u-m_d$
is the only source of IV. This can easily be incorporated in the EFT 
by means of an external scalar source.
Since the charges of the quarks are also different,
there is an additional electromagnetic contribution to IV. 
The electromagnetic effects due to hard photons are represented by local
contact interactions. Soft photons appear in loop diagrams and also
generate the long--range Coulomb potential. In \cite{WME},
these effects have been studied in some detail, extending earlier
work of van Kolck and collaborators \cite{kolck3,KFG}. The power counting
is extended to include the electromagnetic interactions with the
fine structure constant $\alpha$ serving as the additional small parameter.
With this, one can construct the various contributions to the NN potential.
It consists of two distinct pieces, the strong (nuclear) potential 
including isospin violating effects and the Coulomb potential. 
The nuclear potential consists of one-- and two--pion exchange
graphs (with different pion and nucleon masses), $\pi\gamma$ exchange 
diagrams and a set of  local four--nucleon operators (some of which
are isospin symmetric, some depend on the quark charges and some 
on the quark mass difference).  Since the nuclear effective potential
is naturally formulated in momentum space, the matching procedure
developed in ref.\cite{VP} to incorporate the correct asymptotical
Coulomb states was employed in \cite{WME}. More precisely, the
classification of the IV contributions to the NN potential 
is a as follows: To leading  order one has to consider the pion 
mass difference in the OPE and the Coulomb potential. NLO IV
corrections stem from the pion mass
difference in the TPE, from $\pi \gamma$ exchange and from two
four-nucleon contact interactions with no derivatives which have the
generic structure:
\begin{equation}
{\cal O}_{\rm CSB} \sim (N^{\dagger}\tau_{3}N)(N^{\dagger}N),\quad
{\cal O}_{\rm CIB} \sim (N^{\dagger}\tau_{3}N)^{2}~.
\end{equation}
These operators parameterize non-pionic CS breaking and CI breaking effects.
The low--energy constants accompanying the contact interactions
and the cut--off $\Lambda$ were determined in \cite{WME} by a {\em simultaneous}
best fit to the S-- and P--waves of the Nijmegen phase shift analysis 
in the $np$ and the $pp$ systems for
laboratory energies below 50~MeV. This allows to predict these partial
waves at higher energies and all higher partial waves.
Most physical observables come out independent of the sharp cut--off for $\Lambda$
between 300 and 500~MeV. The upper limit on this range is determined by the
$pp$ interactions. 

\begin{figure}[h]
\begin{turn}{270}
\epsfig{file=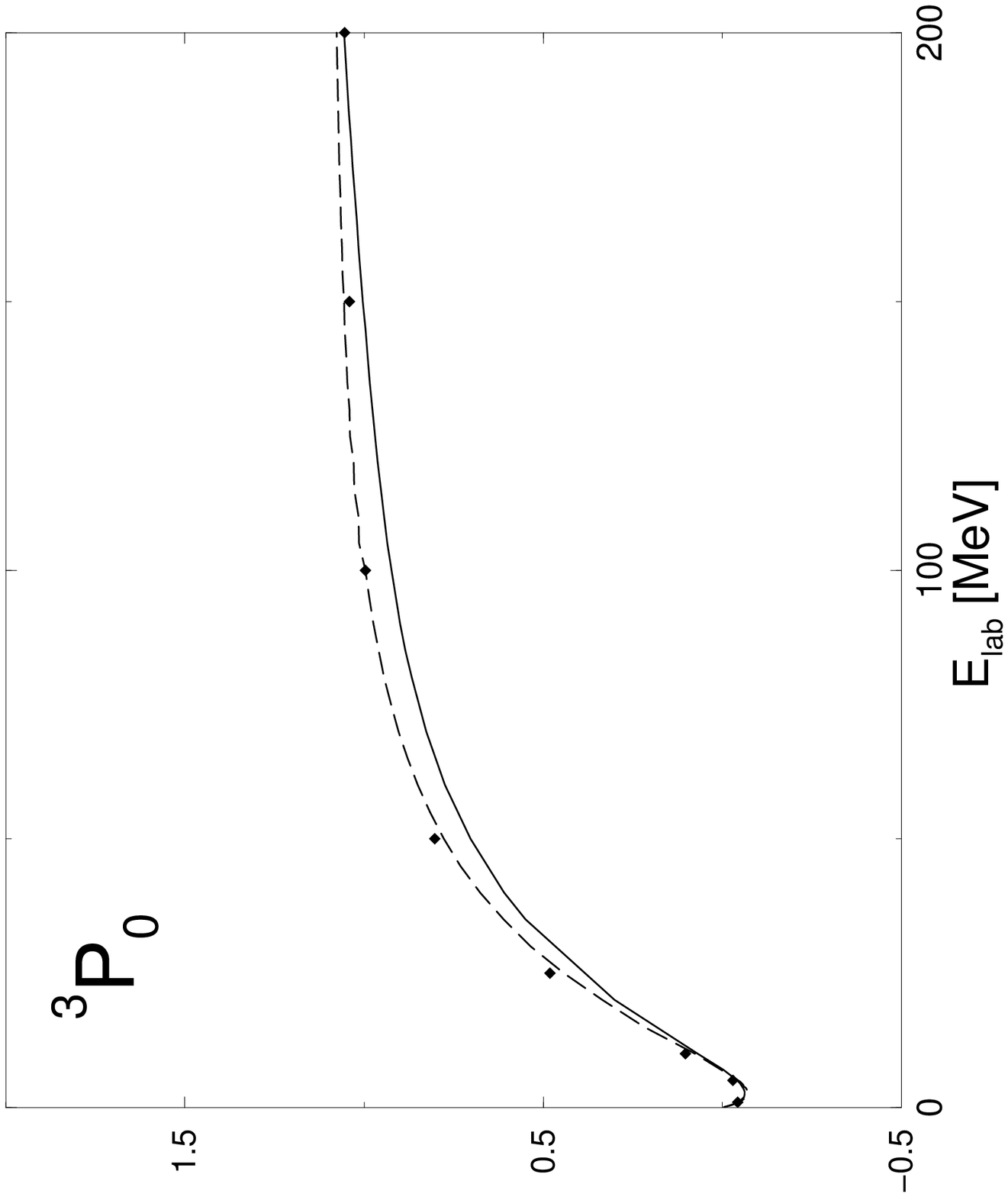, height = 6cm}
\end{turn}
\hspace{1.5cm}
\begin{turn}{270}
\epsfig{file=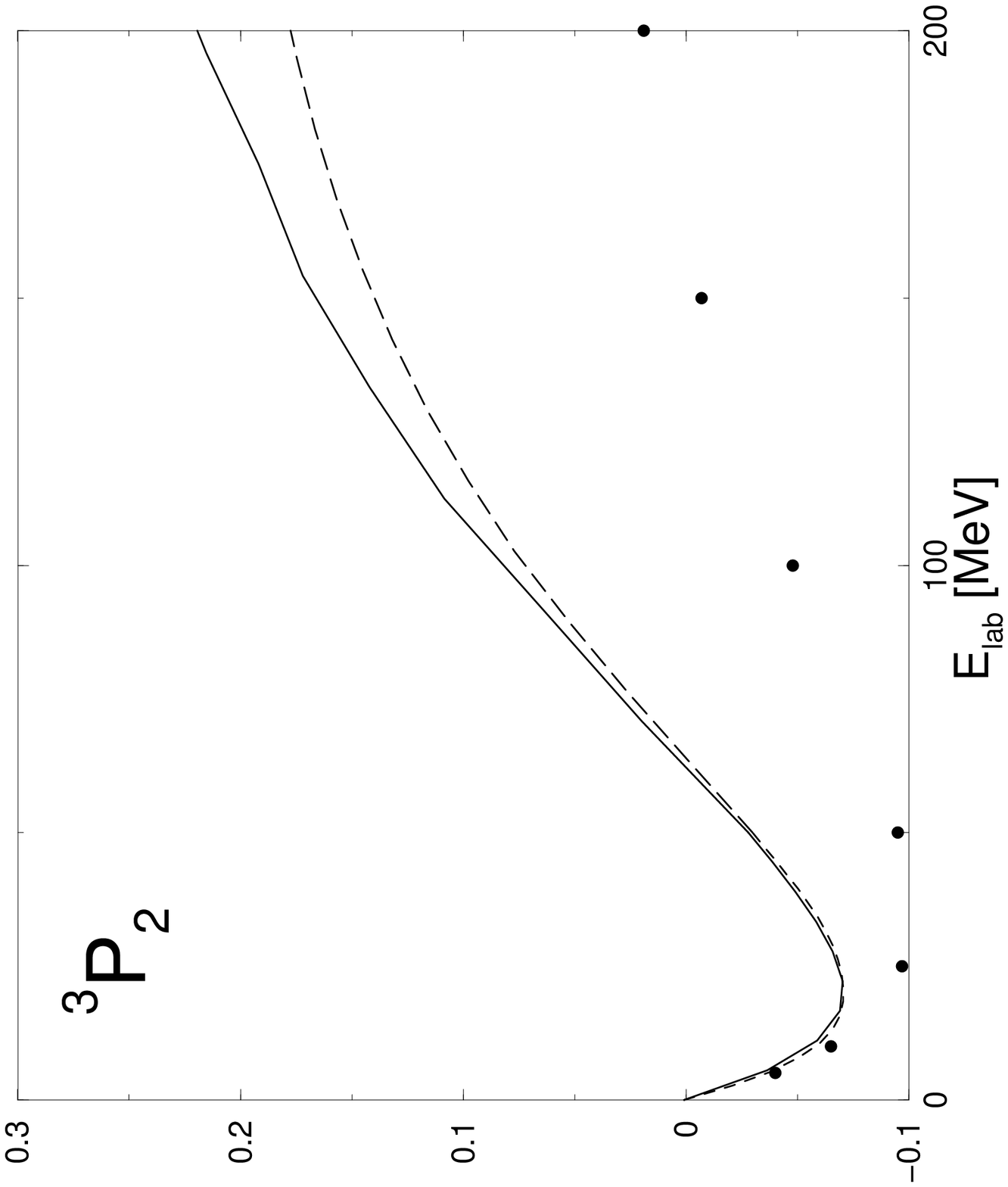, height = 6cm}
\end{turn}
\caption{Phase shift difference $\delta_{\rm pp}- \delta_{\rm
    np}$ for the $^3P_0$ and $^3P_2$ partial waves in comparison to
  the Nijmegen PSA (dashed lines) and results from the 
  CD-Bonn potential (filled circles).}
\label{fig3}
\end{figure}

In fig.\ref{fig3}, we show the predictions for two
P-waves in comparison to the Nijmegen PSA and the CD-Bonn potential.
The range expansion for the $np$ and the $pp$ system was also studied
in \cite{WME}. For the range of cut--offs, the $pp$ scattering length
varies modestly with $\Lambda$  due to the scheme--dependent separation of the nuclear
and the Coulomb part, $|\delta a_{pp} / a_{pp}| \simeq 0.6~{\rm fm}/
17~{\rm fm} \simeq 3\%$. This shows that the arbitrariness in  separating  the Coulomb
and the nuclear part to $a_{pp}$ is strongly reduced within EFT because
only a certain range of cut-offs allows to simultaneously describe
$np$ and $pp$ scattering. Finally, it is worth to stress that CSB is
largely driven by a contact interaction. The corresponding LEC can be understood
to some extent in terms of $\rho-\omega$ mixing, see \cite{KFG}.

\subsection{Three and four nucleons}

\begin{figure}[htb]
\begin{turn}{270}
\psfig{file=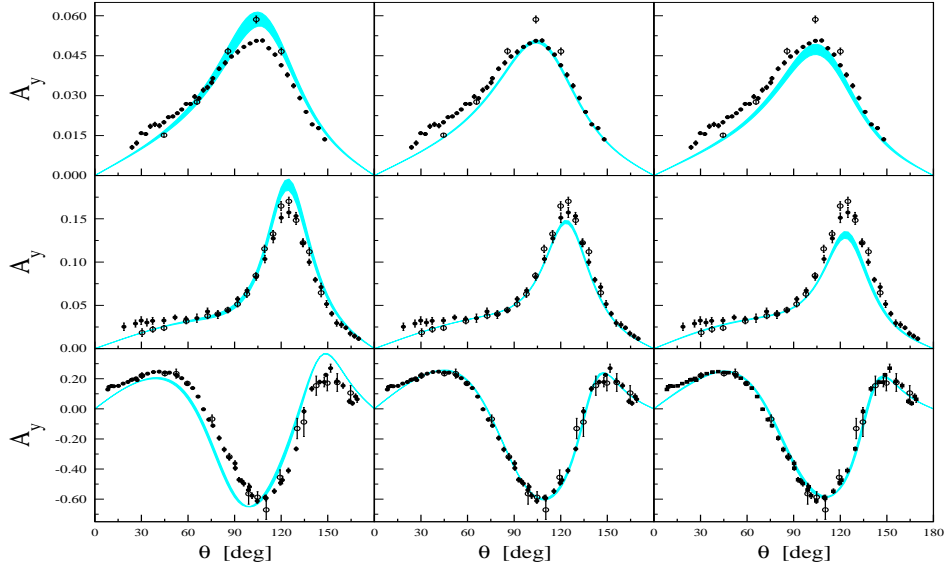,height=5in}
\end{turn}
\vspace{-3cm}
\caption{\label{fig4}Analyzing power $A_y$ for elastic $nd$
scattering, for $E_{\rm lab} = 3,10,65\,$MeV (top to bottom). 
Results at NLO (left panel) and NNLO* (middle
  panel). The band  corresponds to the range $\Lambda = 500$ to $600$ MeV.
Results based on the high-precision potentials (CD-Bonn, AV-18, Nijm-93, Nijm-I,II) 
are shown in the right panel.
Here the band refers to the uncertainty using the various potentials.}
\end{figure}

The NLO $np$ potential was applied  to systems of three and
four nucleons in \cite{ourPRL}. At this order, one has no 3NF 
and thus obtains parameter--free predictions. One also
gets a first indication about the theoretical accuracy of this approach.  
Consider first the binding energy. For changing the 
cut--off between 540 and 600 MeV, the $^3$H and $^4$He binding energies
vary between $-7.55 \ldots -8.28$ and $-24.0 \ldots -28.1$~MeV, 
respectively,
showing the level of accuracy of the NLO approximation for these observables. 
As discussed in detail
in \cite{ourPRL}, the chiral predictions for most observables
like the differential cross section or the tensor analyzing powers $T_{ij}$
for elastic $nd$ scattering as well as break-up observables
are in agreement with what is found for high precision
potentials like CD-Bonn. One also observes a clear improvement 
in the analyzing power $A_y$ for low and moderate energies, 
compare the left and the right panels in fig.\ref{fig4}.
The cut-off dependence for the canonical range of $\Lambda$ is
moderate, as indicated by the band in fig.\ref{fig4}. These
calculations have now been extended to NNLO, more precisely, using the
NNLO* potential and neglecting all 3NFs, which appear at that order.
For a detailed discussion of the results we refer to
\cite{EGMtoappear}. Here, we only remark that the cut-off sensitivity
of the 3N and 4N binding energies is sizeably reduced at NNLO*, one finds e.g.
$-29.96 \ldots -27.87$~MeV for $E_B (^4{\rm He})$ for even extended cut--off range between 500 and 600 MeV
(the overbinding is partly
due to the fact that only an $np$ force is used). Also, the
description of $A_y$ at 3 and 10 MeV is worse than at NLO, but it is still somewhat
better than for the high-precision potentials and, more importantly, 
the cut-off dependence is much weaker as compared to NLO. 
Another crucial observation is that at NNLO* we are able to go to much
higher energies than at NLO: our NNLO* prediction for $A_y$ at 65 MeV is in excellent
agreemnet with the data.
Of course, final conclusions can  only be drawn when the 3NFs have been included. 
Work along these lines is underway. For some first steps in this direction see ref\cite{hueber}.

\subsection{Connection to ``realistic'' potentials}

\begin{figure}[htb]
\psfig{file=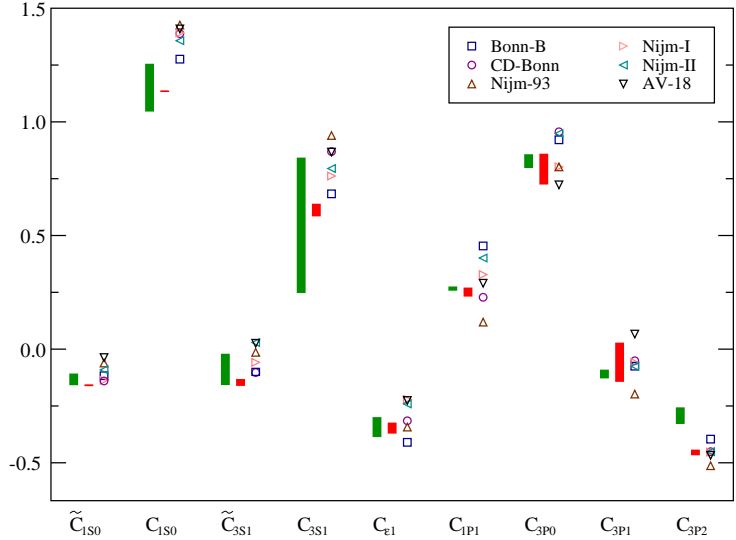,height=2.8in}
\vspace{-.3cm}
\caption{\label{fig6}LECs from phenomenological models and chiral
  EFT. The leftmost band refers to NLO (the length reflects the
  variation with the cut-off), the middle bar is NNLO*, and the
  symbols correspond to the indicated potentials (see inset).
}
\end{figure}

Finally, we wish to provide a bridge between the EFT and traditional nuclear physics
approaches. In the latter case, one constructs (semi)phenomenological potentials such
that one can describe the NN data very precisely. One particular class are the boson
exchange potentials, which besides OPE have heavy meson exchanges ($\sigma$, $\rho$,
$\omega, \ldots$) to generate the intermediate range attraction, short range repulsion
and so on. In most modern high-precision potentials (which lead to fits with a $\chi^2/{\rm datum}
\sim 1$) the short-distance physics is parametized in different ways, say by boundary
conditions, r-space fit functions or partial wave dependent boson exchanges. In \cite{EGME}
it was demonstrated that existing one--boson--exchange
(or more phenomenologically constructed) models of NN
force allow to explain the LECs in the chiral EFT potential in terms of resonance parameters.
To be specific, consider  a heavy meson exchange graph in a generic OBE potential. In the
limit of large meson masses $M_R$, keeping the ratio of coupling constant $g$ to mass
fixed, one can interpret such exchange diagrams as a sum of local operators
with increasing number of derivatives (momentum insertions). 
In a highly symbolic relativistic notation, this reads,
\beq\label{reso}
(\bar N P_i N) \left( {g^2 \, \delta^{ij}\over M_R^2-t} \right) 
(\bar N P_j N)
= \left( {g^2 \over M_R^2} \right) (\bar N P_i N)  (\bar N P^i N)
+ \left( {g^2\, t \over M_R^4} \right) (\bar N P_i N) 
(\bar N P^i N) + \dots~,
\eeq
where the $P_i$ are projectors on the appropriate 
quantum numbers for a given meson exchange (including also Dirac matrices if needed). 
Clearly, it is easy to express the dimension zero and two LECs, 
corresponding to the two terms of the r.h.s. of eq.(\ref{reso}), in terms of meson masses,
coupling constants (and form factor scales, if required by the model). In the chiral EFT, 
one has to expand the TPE in a similar way and adjust its contribution to the various LECs accordingly.
One can also repeat this procedure for the high-precision potentials, this has to 
be done numerically for the various partial waves. In fig.\ref{fig6} we show a comparison
between the 4 S- and the 5 P-wave LECs obtained at NLO (leftmost bands) and NNLO* (central bands) and
extracted from various OBE and high-precision potentials (see the inset in the figure). The agreement
is rather stunning since one would have expected higher dimension operators to play
a more prominent role in the phenomenological potentials (related  e.g. to the cut-off sensitivity
of the $\pi N$ form factor like in the Bonn potential). Note also
that the theoretical uncertainties determined in EFT are a) small compared to their
average values and b) are smaller than the band spanned by the potential models (even
if one only includes the high--precision ones). Note that the comparison has been performed for the 
partial wave projected set of coupling constants.
\noindent
Let us now discuss the important  issue of  naturalness of the various LECs related to contact interactions.
For that is turns out to be more appropriate to work directly with the coupling constants $C_S$,
$C_T$, $C_{1\ldots 7}$,
which enter the chiral Lagrangian, see \cite{EGME}. Here, the constants $C_S$ and $C_T$ correspond 
to operators without derivatives, whereas the remaining ones to contact terms with two derivatives.
Dimensional arguments suggest the following scaling properties of these LECs:
$C_{C,S} \sim {l_{S,C} / (f_\pi^2) }\;$,  $C_{1 \ldots 7} \sim {l_{1 \ldots 7} / (f_\pi^2 \, \Lambda_\chi^2)}$,
where the $l$'s are some numbers of order one. 
These scaling properties of the LECs are crucial for the convergence of the low--momentum expansion. 
Note that one has to take into account 
numerical prefactors which accompany the various terms
of the Lagrangian, see ref.~\cite{EGME} for more details.
Taking $\Lambda_\chi =1$ GeV it turnes out that 
the values of the $l$'s fluctuate
between 0.3 and 3.5, i.e. the values found for these LECs are
indeed natural, with the notable exception of $f_\pi^2 \, C_T$,
which  is much smaller than one: $f_\pi^2 C_T = -0.002 \ldots 0.147$ at NLO and 
$f_\pi^2 C_T = 0.002 \ldots 0.040$ at NNLO*. These unnaturally small numbers can be
traced back to the SU(4) symmetry of the NN interaction, proposed about 65 years ago by Wigner 
\cite{Wigner}. For the recent discussion on that subject within the EFT approach see ref.~ \cite{MSW}.

\section{Summary and outlook}

Despite the original scepticism by its creator \cite{Weinb2}, chiral
effective field theory not only offers qualitative but also {\em
  quantitative} insight into the dynamics of few-nucleon systems, as
should have become clear from the discussed topics. In addition, it is
the only framework at present in which one can address the question of the size
of three (four) nucleon forces in a truely systematic manner. It is
therefore evident that a vast effort has to be undertaken to pin down
the chiral 3NF and apply it not only to few- but also many-body
systems. Exciting times are ahead of us.

\section{Acknowledgments}
E.E. thanks the organizers for the invitation and support.

\end{document}